\documentclass{PoS}
\usepackage{graphicx}
\usepackage{amssymb}
\usepackage{epstopdf}
\usepackage{slashbox}

\newcommand {\snn}      {\sqrt{s_{_{\rm NN}}}}
\title{Measurement of directed flow of $D^{0}$ and $\bar{D^{0}}$ mesons in 200 GeV Au+Au collisions at RHIC using the STAR detector}

\ShortTitle{Short Title for header}

\author{\speaker{Liang He (for the STAR collaboration)}\\
        Purdue University\\
        E-mail: \email{he202@purdue.edu}}

\abstract{

Charm quarks, owing to their large mass, are produced predominantly in the initial hard scatterings in heavy-ion collisions, and therefore can be a valuable tool for studying the early time dynamics of these collisions. The rapidity-odd directed flow at mid-rapidity in heavy-ion collisions originates from a tilt in the reaction plane of the  thermalized medium caused by the asymmetry between the number of participants from projectile and target nuclei as a function of rapidity. 
Recently, it has been predicted that the slope of the directed flow at mid-rapidity for $D^0$ mesons, arising from the transport of charm quarks in the tilted medium, can be several times larger than that of light flavor hadrons. 
The magnitude of the slope is expected to be sensitive to the magnitude of the tilt and the charm quark drag coefficient in the medium. It has also been predicted that the transient electromagnetic field generated at early time can induce a much larger directed flow for heavy quarks than for 
light quarks.
In these proceedings, we will report on the first measurement of the directed flow for the $D^0$ and $\bar{D}^0$ mesons as a function of rapidity in Au+Au collisions at $\snn$ = 200 GeV using high statistics data collected with the Heavy Flavor Tracker in 2014 and 2016 RHIC runs. 
          }

\FullConference{International Conference on Hard and Electromagnetic Probes of High-Energy Nuclear Collisions\\
		30 September - 5 October 2018\\
		Aix-Les-Bains, Savoie, France}

\begin{document}
\section{Introduction}
Heavy quarks are predominantly produced in hard scatterings at the early stage of relativistic heavy-ion collisions. They subsequently experience the entire evolution of the collision system including the quark-gluon plasma (QGP) phase. 
Moreover, the relaxation time of heavy quarks is comparable to the lifetime of the QGP, making them a unique probe of the early time dynamics of the QGP \cite{Rapp:2008qc}. 

The directed flow ($v_1$), defined as the first harmonic coefficient of the particle azimuthal distribution, is also believed to be a sensitive probe into the early time dynamics of heavy-ion collisions \cite{v1:1992, v1:1998, v1:2011}. 
A hydrodynamic calculation with a tilted initial source \cite{hydro} can explain the negative $v_1$-slope or "anti-flow" of charged hadrons measured at RHIC energies at midrapidity \cite{anti}. Recently a framework, based on Langevin dynamics for heavy quarks coupled to the hydrodynamic background calculation, predicts a stronger $v_1$ for the $D$ mesons compared to the light hadrons \cite{langevin}. 
The measurement of the $D$ meson $v_1$ can hence be used to constrain the drag coefficients of the tilted bulk. 
Furthermore, a larger $v_1$ for heavy quarks is predicted to result from the transient magnetic field generated in heavy-ion collisions. The signs of $v_{1}$, induced by the initial electromagnetic field, are predicted to be opposite for charm and anti-charm quarks, although the magnitude of the resulting $v_1$ splitting may be smaller than the overall $v_1$ induced by the drag from the tilted source \cite{em1, em2}. 
The $v_1$ splitting for $D^0$ and $\bar{D}^0$ may provide insights into the early-time electromagnetic field generated in heavy-ion collisions.
In these proceedings, we report on the measurements of $D^0$ and $\bar{D}^0$ $v_1$ in Au+Au collisions at $\snn$ = 200 GeV at the STAR \cite{star}  experiment.

\section{Data analysis}
In this analysis $D^{0}$ mesons are reconstructed through the hadronic decay channel $D^{0}(\bar{D^{0}})\rightarrow K^{-}\pi^{+}(K^{+}\pi^{-})$, with $c\tau \approx$ 122.9 $\mu$m and branching ratio of $3.89 \pm 0.04\%$. 

Charged tracks are reconstructed with the Time Projection Chamber (TPC) \cite{tpc} and the Heavy Flavor Tracker (HFT) \cite{hft}. The TPC covers the full azimuth at mid-rapidity ($-1<\eta<1$). The HFT, a high resolution silicon detector which enables the precise topological reconstruction of the charmed hadrons, has a similar coverage as the TPC. To ensure good HFT acceptance, the reconstructed primary vertex is required to be within 6 cm along the z-direction from the center of the TPC.
STAR \cite{star} collected about 2 billion events with minimum bias (MB) trigger events with the HFT in 2014 and 2016. 

The particle identification is done with the TPC and the Time of Flight (TOF) detector \cite{tof}. The TPC provides energy loss measurements ($dE/dx$), which are required to be within 2 (for $K^{\pm}$) or 3 (for $\pi^{\pm}$) standard deviations from the expected mean value. If the TOF information is available, the inverse velocity $1/\beta$ is required to be within 0.03 from the expected mean value. 
The decay vertex is reconstructed as the mid-point of the distance of the closest approach between the tracks of $K^{\pm}$ and $\pi^{\mp}$. 
Cuts on the decay topology enabled by the HFT are used to suppress combinatorial background \cite{reco}. 

\begin{figure}[h]
\centering
\begin{minipage}{14pc}
\includegraphics[width=14pc]{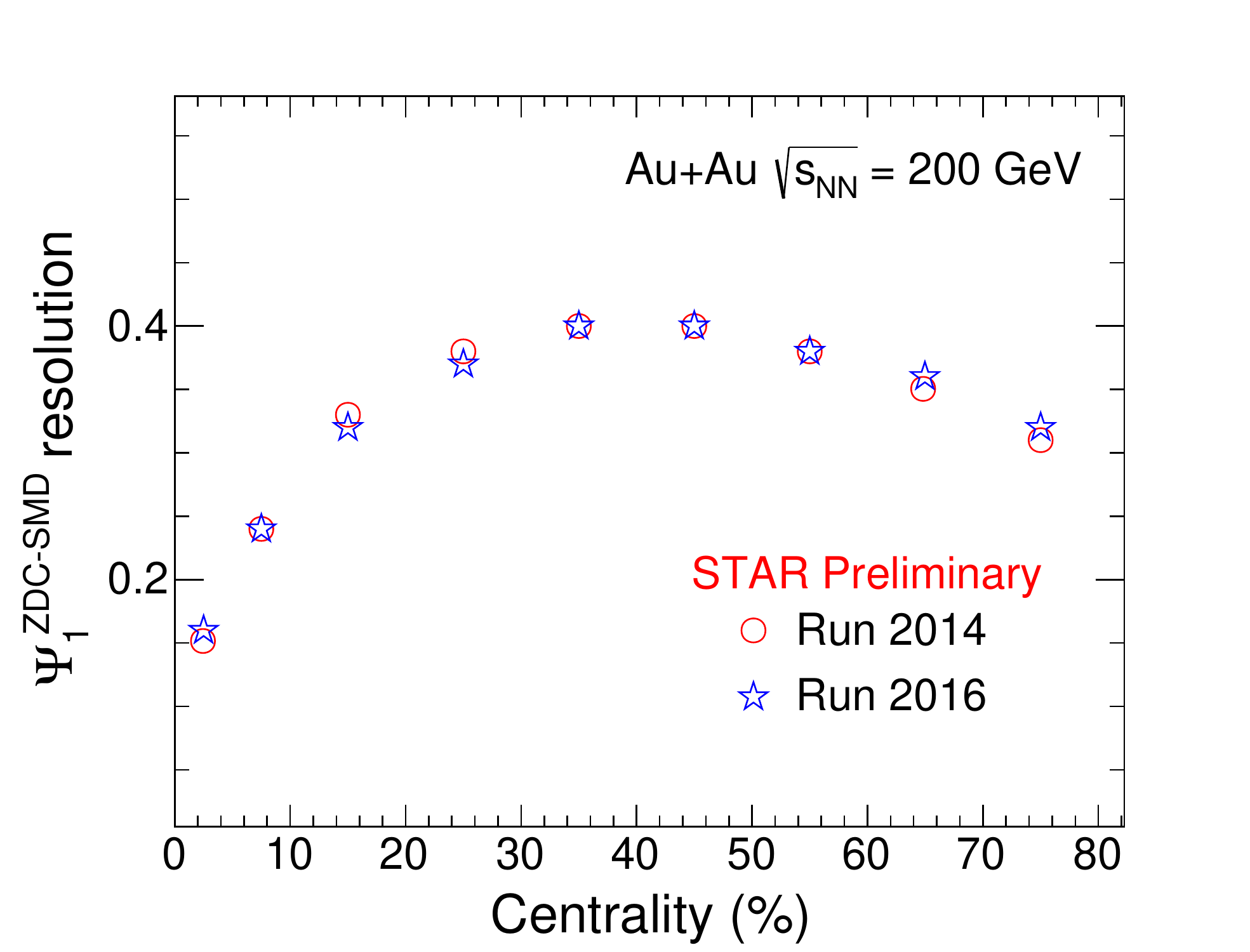}
\end{minipage}\hspace{2pc}%
\begin{minipage}{14pc}                                                    
  \includegraphics[width=14pc]{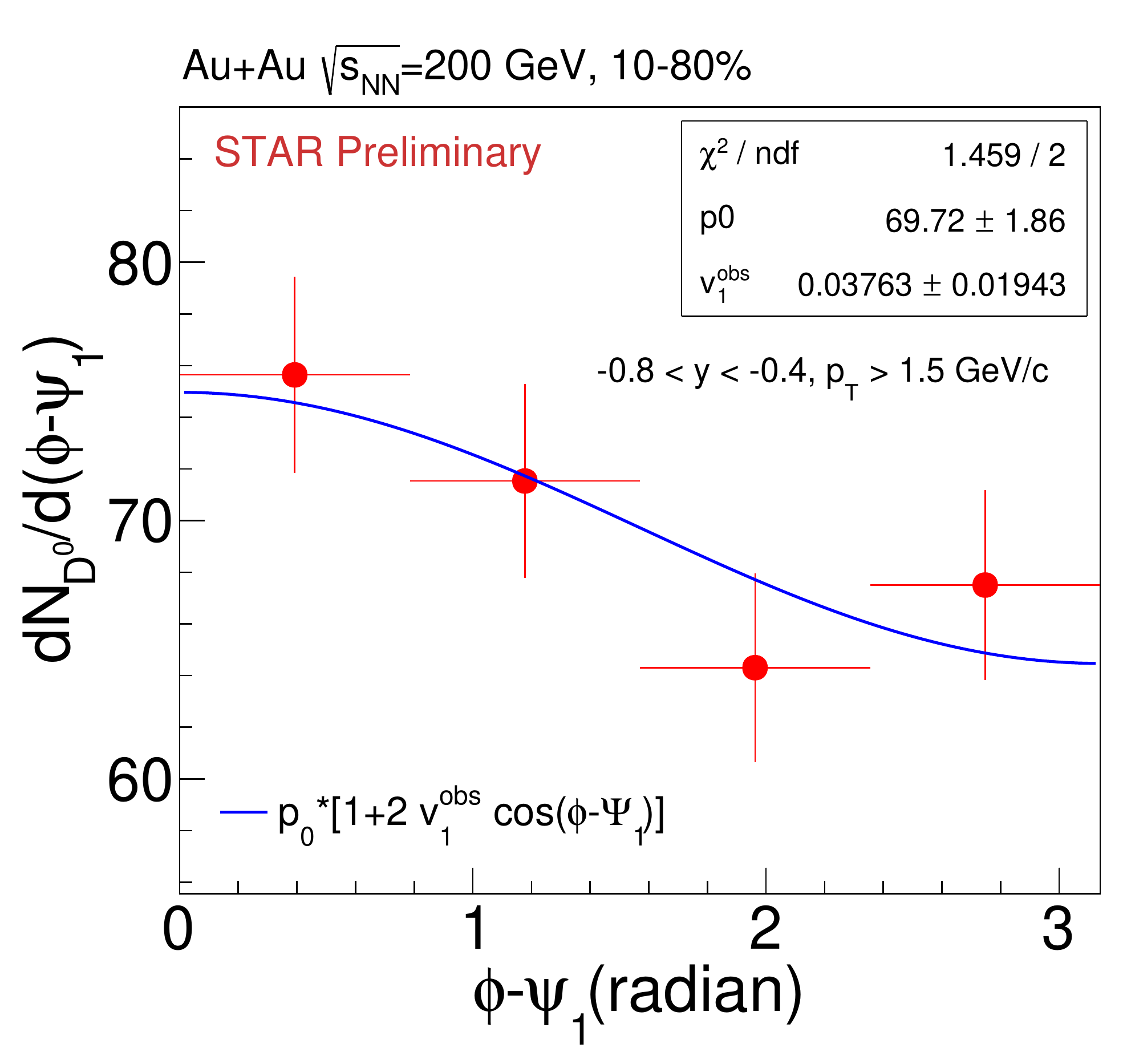}
\end{minipage}
  \caption{Left:\label{fig:resolution} The ZDC-SMD first-order event plane resolution as a function of centrality. The red circle and blue star markers represent the 2014 and 2016 data respectively;
  Right:\label{fig:phipsi} An example of fitting $D^0$ yield as a function of $\phi-\psi_1$ for $p_{T}(D^{0}) > 1.5$ GeV/$c$ within $-0.8<y< -0.4$.} 
\end{figure}

The first order event plane ($\psi_1$) is reconstructed from the east and west Zero Degree Calorimeter Shower Maximum Detectors (ZDC-SMD) \cite{ep}. The ZDC-SMD is located at $|\eta|$ $>$ 6.3. Due to five units of $\eta$-gap of ZDC-SMDs from the TPC and HFT, non-flow effects are significantly reduced \cite{nonflow}. The first order event plane resolution \cite{reso} is shown in the left panel of Figure~\ref{fig:resolution}.

The $D^0$ $v_1$ is calculated using the event plane method \cite{nonflow}. 
The $D^0$ yield is obtained by integrating the invariant mass distribution of unlike-sign $K\pi$ pairs within a window of 1.82 - 1.91 GeV/$c^2$ and subtracting the combinatorial background. 
The yield is obtained in each $\phi - \psi_1$ bin in four different rapidity windows, and is weighted by the inverse of the detector efficiency times the acceptance. The observed $v_1$ is calculated by fitting the yield with a function, $p_{0}[1+2v_{1}\cos(\phi-\psi_1)]$, as shown in the right panel of Figure~\ref{fig:phipsi}. The final $v_1$ is obtained by scaling the $v_{1}^{obs}$ with $\langle 1/R \rangle$, where $R$ is the event plane resolution \cite{reso}.

\section{Results}
\begin{figure}[h]
\centering
\includegraphics[width=14pc]{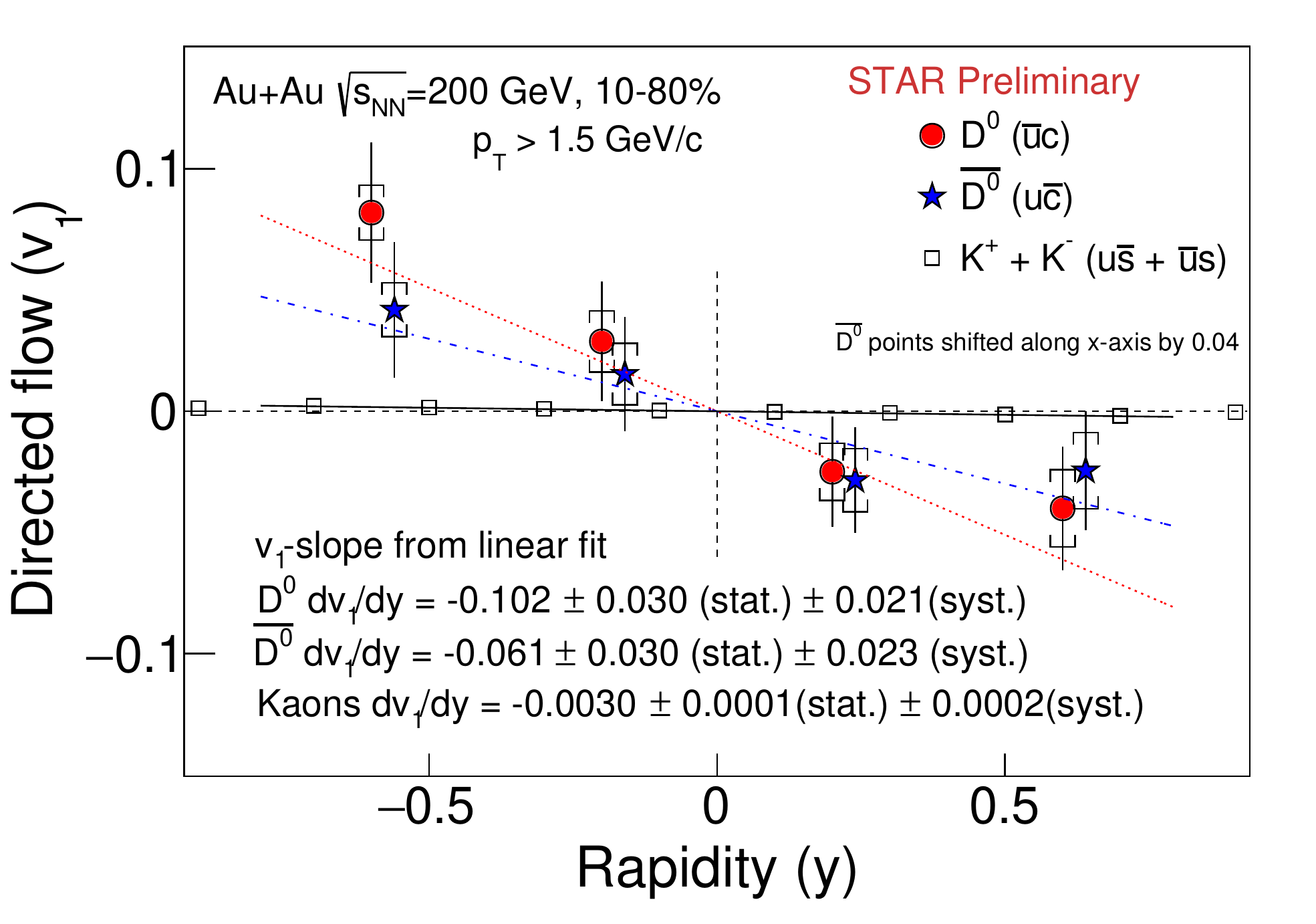}
\caption{
    \label{fig:kaon} Red circles and blue stars represent $D^0$ and $\bar{D}^{0}$ $v_1$ as a function of rapidity for $p_{T}$ $>$ 1.5 GeV/$c$ in 10-80$\%$ central Au+Au collisions at $\snn$ = 200 GeV.  The open squares represent the average $v_1$ for charged kaons. The $D^0$ and $\bar{D}^0$ $v_1$ are fit with a linear function and plotted as red and blue lines.
  }
\end{figure}

Figure~\ref{fig:kaon} shows the measured rapidity dependence of $v_1$ for the $D^0$ and $\bar{D}^0$ mesons in 10-80$\%$ Au+Au collisions at $\snn$ = 200 GeV. For clarity, the $\bar{D}^0$ points are shifted along the x-axis by 0.04. The $v_1$-slope ($dv_{1}/dy$) is obtained by fitting the data with a linear function passing through zero. The $D^0$ $v_1$ results are compared to charged kaons (open square markers). 
The $dv_{1}/dy$ for combined $D^0$ and $\bar{D}^0$ is $-0.08 \pm 0.02$(stat.)$\pm 0.02 $(syst.), which is about 20 times larger than that of kaons with a 3 $\sigma$ level of significance.

\begin{figure}[h]
\centering
\begin{minipage}{14pc}
  \includegraphics[width=14pc]{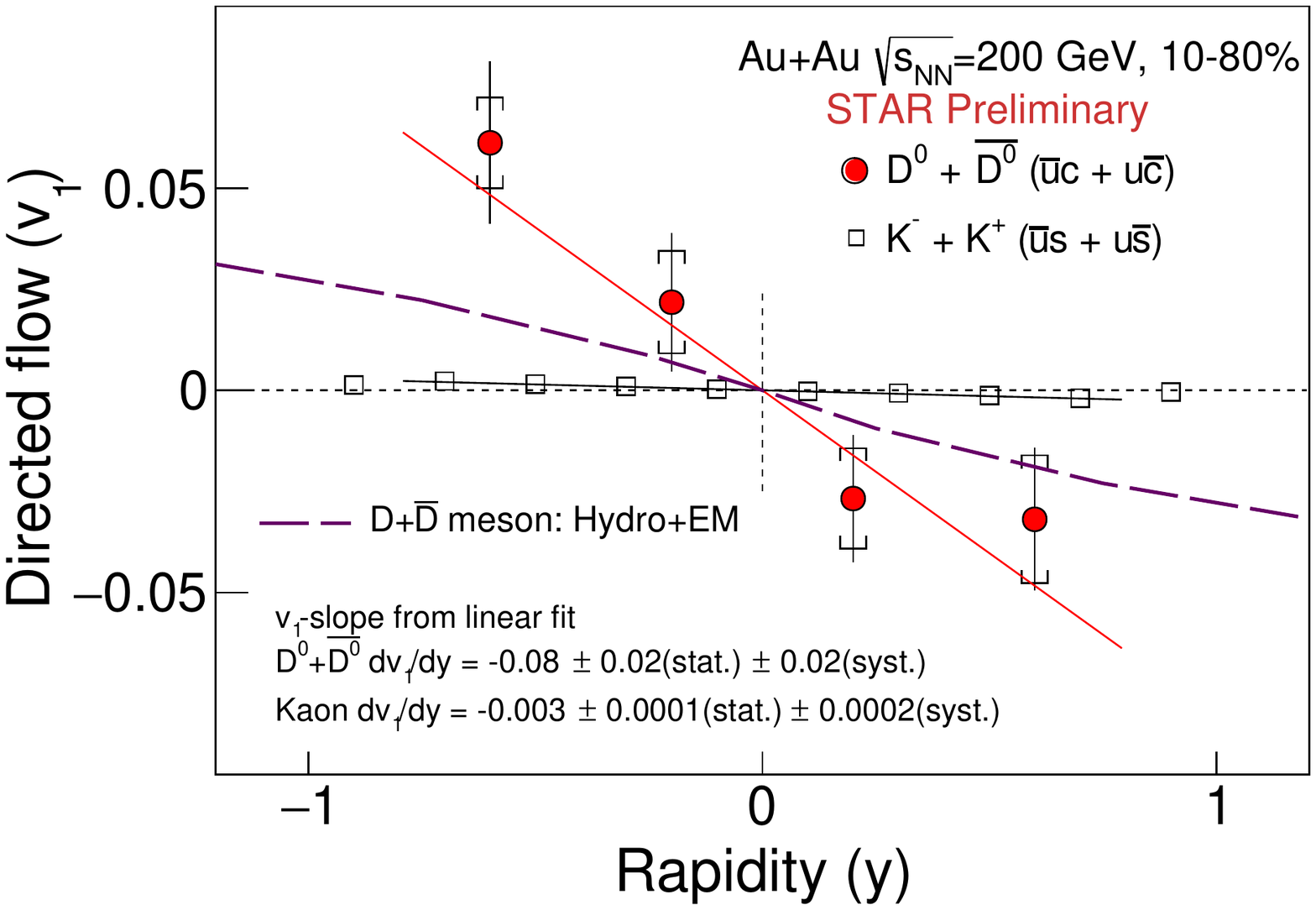}
\end{minipage}\hspace{2pc}%
\begin{minipage}{14pc}                                                    
  \includegraphics[width=15pc]{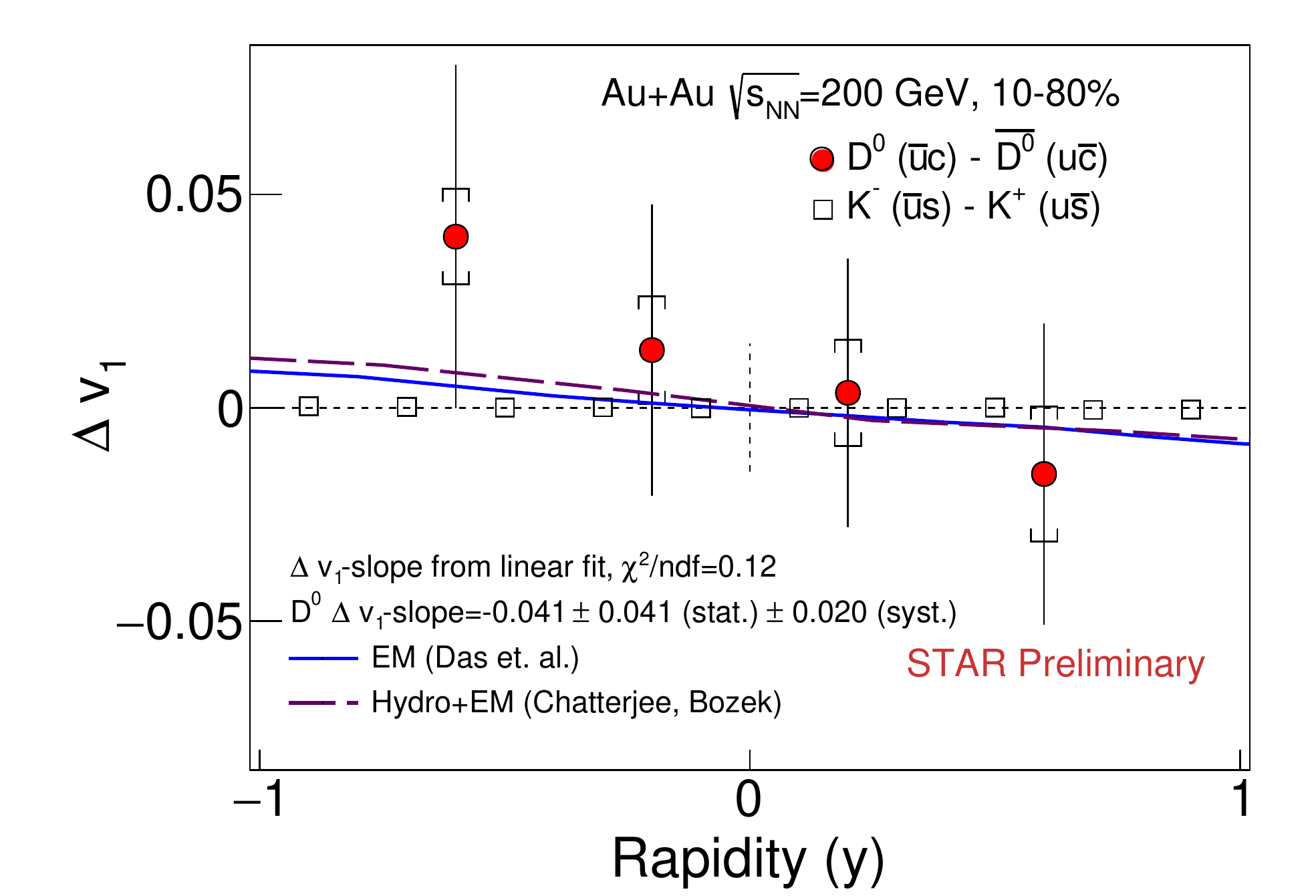}
\end{minipage}
  \caption{
    Left:\label{fig:model} The red circles represent the average $v_1$ for combined $D^0$ and $\bar{D}^0$ for $p_T$ $>$ 1.5 GeV/$c$ in 10-80$\%$ Au+Au collisions at $\snn$ = 200 GeV. The open squares are for average $v_1$ of charged kaons. The magenta dashed line is a hydrodynamic model calculation combined with the initial electromagnetic field;
  Right:\label{fig:delta} The red circles represent the difference in $v_1$ ($\Delta v_{1}$) between $D^0$ and $\bar{D}^0$. The open squares are for $\Delta v_1$ between $K^-$ and $K^+$ mesons. The blue solid line and magenta dashed line are $D$ meson $\Delta v_1$ predictions from the initial electromagnetic field only and from hydrodynamics combined with the initial electromagnetic field respectively. 
  }
\end{figure}

The $D^0$ $v_1$ results are also compared to a hydrodynamic model in the left panel of Figure~\ref{fig:model}. The model calculation, combining the Langevin dynamics for the heavy quarks within the hydrodynamical background from the tilt bulk together with the initial electromagnetic field \cite{langevin}, predicts the correct sign of $dv_{1}/dy$ for both $D^0$ and $\bar{D}^0$, but underestimates the magnitude with the choice of the used model parameters. Our results can therefore help constrain the model parameters, such as the tilt and charm quark drag coefficients. 

The right panel of Figure~\ref{fig:delta} shows the comparison of the measured difference in $v_1$ ($\Delta v_1$) between $D^0$ and $\bar{D}^0$ to the theoretical predictions. The dashed magenta line is the calculation from a hydrodynamic model incorporating both the tilted bulk and the initial electromagnetic field \cite{langevin}. The solid blue line is from the initial electromagnetic field only \cite{em1}. The measured $\Delta v_1$-slope is $-0.041\pm0.041$(stat.)$\pm 0.020$(syst.), which is consistent with zero as well as with model calculations. The current precision of the data is not sufficient to draw conclusions regarding the effect of electromagnetic field.

\section{Conclusions}
In summary, we report on the measurement of the rapidity-odd directed flow ($v_{1}(y)$) for $D^0$ and $\bar{D}^0$ mesons in 10-80$\%$ central Au+Au collisions at $\snn$ = 200 GeV using the STAR detector at RHIC. 
The measured $v_1$-slope is about 20 times larger than that of kaon with a 3 $\sigma$ significance. It indicates strong interaction of charm quarks with the initially tilted source. The negative $dv_{1}/dy$ slopes for $D^0$ and $\bar{D}^0$ are observed as predicted by theoretical calculation. The current measurement precision is not sufficient to draw firm conclusions about the splitting between  $D^0$ and $\bar{D}^0$ $v_1$ induced by the initial electromagnetic field.


\begin{thebibliography}{99}
\bibitem{Rapp:2008qc}
  R. Rapp and H. van Hees, {arXiv: 0803.0901} (2008).
\bibitem{v1:1992}
  J.-Y. Ollitrault, Phys. Rev. \textbf{D46}, 229 (1992).
\bibitem{v1:1998}
  A. M. Poskanzer and S. A. Voloshin, Phys. Rev. \textbf{C58}, 1671 (1998), nucl-ex/9805001.
\bibitem{v1:2011}
  A. Bilandzic, R. Snellings, and S. Voloshin, Phys. Rev. \textbf{C83}, 044913 (2011), 1010.0233.
\bibitem{hydro}
  P. Bozek and I. Wyskiel, Phys. Rev. \textbf{C81}, 054902 (2010), 1002.4999.
\bibitem{anti}
  J. Brachmann, S. Soff, A. Dumitru, H. Stoecker, J. A. Maruhn, W. Greiner, L. V. Bravina, and D. H. Rischke, Phys. Rev. \textbf{C61}, 024909 (2000), nucl-th/9908010.
\bibitem{langevin}
    S. Chatterjee and P. Bozek, arxiv:1804.04893
\bibitem{em1}
  S. K. Das, S. Plumari, S. Chatterjee, J. Alam, F. Scardina, and V. Greco, Phys. Lett. \textbf{B768}, 260 (2017), 1608.02231.
\bibitem{em2}
  U. Gursoy, D. Kharzeev, and K. Rajagopal, Phys. Rev. \textbf{C89}, 054905 (2014), 1401.3805.
\bibitem{star}
  K. H. Ackermann et al. (STAR), Nucl. Instrum. Meth. \textbf{A499}, 624 (2003).
\bibitem{tpc}
  M. Anderson et al., Nucl. Instrum. Meth. \textbf{A499}, 659 (2003), nucl-ex/0301015.
\bibitem{hft}
  G. Contin, L. Greiner, et al., Nucl. Inst. Meth. A 2018, \textbf{907}, Pgs. 60-80
\bibitem{tof}
  B. Bonner, H. Chen, G. Eppley, F. Geurts, J. Lamas Valverde, C. Li, W. J. Llope, T. Nussbaum, E. Platner, and J. Roberts, Nucl. Instrum. Meth. \textbf{A508}, 181 (2003).
\bibitem{reco}
  L. Adamczyk \textit{et al.} (STAR), Phys. Rev. Lett. \textbf{118}, 212301 (2017), 1701.06060.
\bibitem{ep}
  J. Adams et al. (STAR), Phys. Rev. \textbf{C73}, 034903 (2006), nucl-ex/0510053.
\bibitem{nonflow}
  A. M. Poskanzer and S. A. Voloshin, Phys. Rev. \textbf{C58}, 1671 (1998), nucl-ex/9805001.
\bibitem{reso}
  H. Masui, A. Schmah, and A. M. Poskanzer, Nucl. Instrum. Meth. \textbf{A833}, 181 (2016), 1212.3650.
 
\end{thebibliography}
\end{document}